\documentclass[prl,superscriptaddress,floatfix,twocolumn]{revtex4-1}
\usepackage{graphicx}
\usepackage{wasysym}
\usepackage{color}

\def\lsco{La$_{2-x}$Sr$_x$CuO$_4$}
\def\lbco{La$_{2-x}$Ba$_x$CuO$_4$}

\def\ybco{YBa$_2$Cu$_3$O$_{6+x}$}

\newfont{\fc}{cmssbx10 scaled 1000}

\begin{document}

\title{Neutron Scattering Evidence for a Periodically-Modulated Superconducting Phase in the Underdoped Cuprate La$_{1.905}$Ba$_{0.095}$CuO$_4$}

\author{Zhijun Xu}
\altaffiliation[Present address: ]{Lawrence Berkeley National Lab, 1 Cyclotron Road, Berkeley, CA 94720}
\affiliation{Condensed Matter Physics \&\ Materials Science Department, Brookhaven National Laboratory, Upton, NY 11973-5000, USA}
\author{C. Stock}
\altaffiliation[Present address: ]{School of Physics \&\ Astronomy, The University of Edinbugh, James Clerk Maxwell Building, Mayfield Rd., Edinburgh EH9 3JZ, United Kingdom}
\author{Songxue Chi}
\altaffiliation[Present address: ]{Oak Ridge National Laboratory, Oak Ridge, Tennessee 37831, USA}
\affiliation{NIST Center for Neutron Research, National Institute of
Standards and Technology, Gaithersburg, Maryland 20899, USA}
\author{A. I. Kolesnikov}
\affiliation{Chemical and Engineering Materials Division, Oak Ridge National Laboratory, Oak Ridge, Tennessee 37831, USA}
\author{Guangyong Xu}
\author{Genda Gu}
\author{J. M. Tranquada}
\affiliation{Condensed Matter Physics \&\ Materials Science Department, Brookhaven National Laboratory, Upton, NY 11973-5000, USA}
\date{\today}
\begin{abstract}
The role of antiferromagnetic spin correlations in high-temperature superconductors remains a matter of debate.  We present inelastic neutron scattering evidence that gapless spin fluctuations coexist with superconductivity in La$_{1.905}$Ba$_{0.095}$CuO$_4$.  Furthermore, we observe that both the low-energy magnetic spectral weight and the spin incommensurability are enhanced with the onset of superconducting correlations.  We propose that the coexistence occurs through intertwining of spatial modulations of the pair wave function and the antiferromagnetic correlations.  This proposal is also directly relevant to sufficiently underdoped \lsco\ and \ybco.
\end{abstract}
\maketitle

It is commonly accepted that cuprate superconductors have a spatially-uniform $d$-wave pair wave function \cite{tsue00}.  It has also become a paradigm that antiferromagnetic spin fluctuations are gapped in the superconducting state, with a pile up of excitations in the magnetic ``resonance'' peak above the gap \cite{scal12a,esch06,yu09,ross91,mook93}.  A number of neutron scattering studies of underdoped \lsco\ have found evidence for incommensurate spin fluctuations that remain gapless at temperatures far below the superconducting transition temperature, $T_c$ \cite{lee00,chan07,lips09,kofu09}.  Theoretical analyses have tended to view such spin-density-wave correlations as soft fluctuations of an order that competes with spatially-uniform superconductivity \cite{deml01} and that may be locally pinned by disorder \cite{ande10}.  As a consequence, researchers have crafted interpretations of the low-energy spin fluctuations that maintain consistency with the spin-gap paradigm \cite{chan07,kofu09}.

In an alternative approach, the superconductivity and antiferromagnetism are both treated as spatially modulated and intimately intertwined \cite{berg09b}.  Such a state, which variational calculations indicate to be energetically competitive with uniform superconductivity \cite{corb14}, has been invoked \cite{hime02,berg07} to explain the depression of superconducting order in certain stripe-ordered cuprates \cite{taji01,li07}.  While the poorly-superconducting phase is fascinating on its own, it leaves open the question of whether a modulated pair wave function might be relevant to the case of a good bulk superconductor.

In this paper, we present neutron scattering measurements of the low-energy spin fluctuations in La$_{1.905}$Ba$_{0.095}$CuO$_4$, a bulk superconductor with $T_c=32$~K.  Rather than developing a spin gap on cooling below $T_c$, the lowest-energy excitations are actually enhanced.  By putting the measurements on an absolute scale, we show that the strength of the spin response is comparable to that of spin waves in antiferromagnetic La$_2$CuO$_4$.   To generate this large a response, we conclude that all parts of the sample must contribute to the signal, ruling out macroscopic phase separation.  A previous optical conductivity study has shown that the superfluid density of this sample is consistent with the trend established for bulk superconductivity in all cuprate families in the form of Homes' law \cite{home12}.  It thus appears that there must be local coexistence of the spin fluctuations and superconductivity.  This view is supported by changes in the low-energy magnetic spectral weight and incommensurability that correlate with the onset of superconductivity.  Given the empirical observation that commensurate antiferromagnetism and superconductivity do not coexist, at least in single-layer cuprates \cite{birg06}, the best option to reconcile the new results is to have a superconducting state that is spatially modulated to minimize overlap with the amplitude-modulated antiferromagnetic correlations.

The single crystal of La$_{1.905}$Ba$_{0.095}$CuO$_{4}$ used here, a
cylinder of size $\diameter 8\mbox{\ mm}\times35$~mm and mass $\sim 11$~g, was grown by the floating-zone technique at Brookhaven \cite{huck11}.  Previous neutron scattering measurements provided evidence for weak charge and spin stripe order \cite{wen12}.  The signal is averaged over the sample, so one cannot distinguish between uniformly weak order and macroscopic phase separation, such as occurs in oxygen-doped \lsco\ \cite{udby13}.   Here we focus on the spin fluctuations in order to deduce the bulk behavior.

\begin{figure*}[t]
\centerline{\includegraphics[width=5.in]{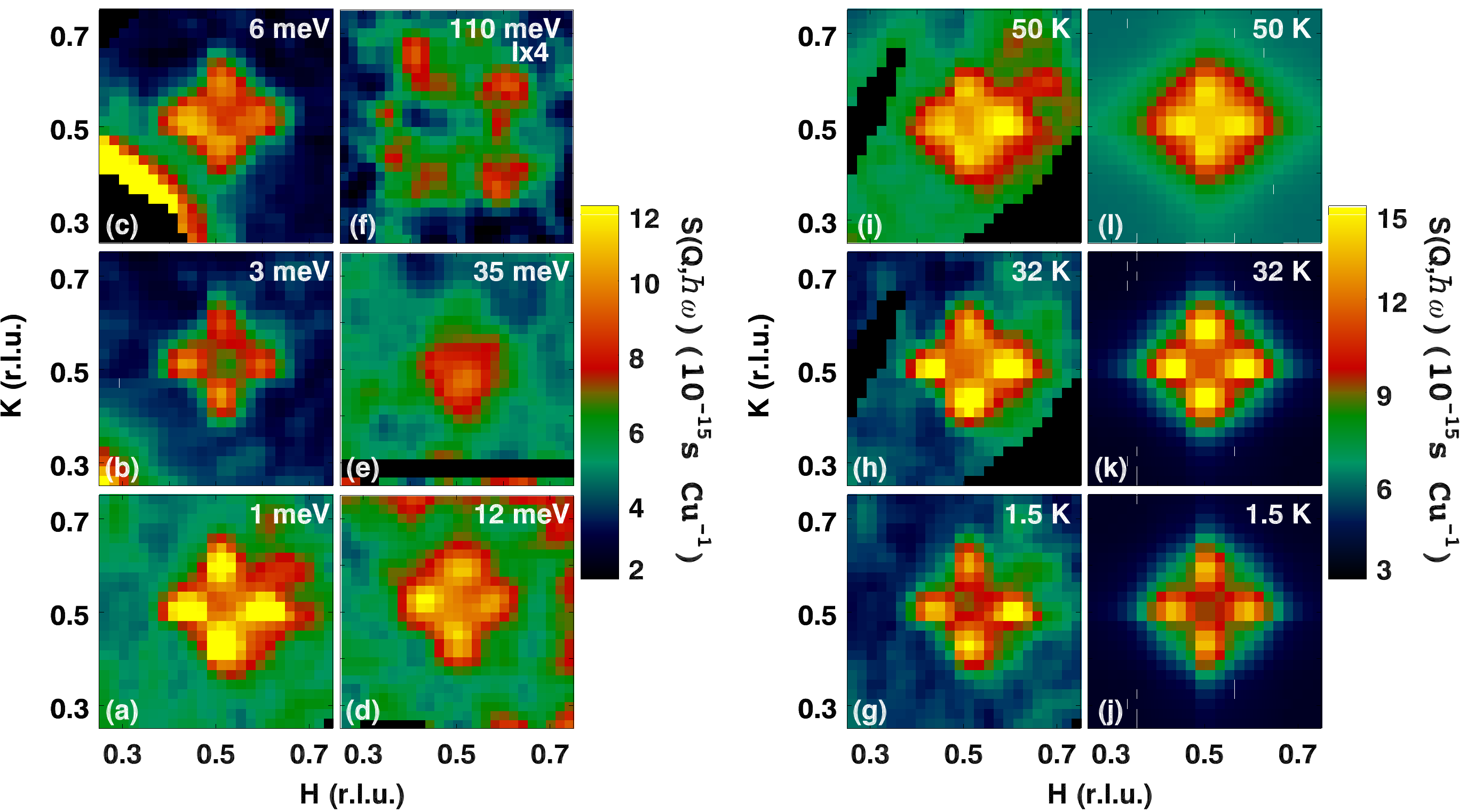}}
\caption{(color online)
Constant-energy slices through $S({\bf Q},\omega)$.  On the left are measurements at $T\lesssim5$~K for excitation energies (a) 1~meV, (b) 3~meV, (c) 6~meV, (d) 12 meV, (e) 35~meV, (f) 110~meV.  On the right are measurements at $\hbar\omega=1$~meV for temperatures of (g) 1.5~K, (h) 32~K, (i) 50~K, together with corresponding fits of symmetrically-positioned Gaussian peaks in (j)--(l).   Data in (a)--(c) and (g)--(i) were obtained at the MACS spectrometer at NCNR; data in (d)--(f) were measured at SEQUOIA (SNS).   The units of $S({\bf Q},\omega)$ are determined by Eq.~(\ref{eq:S}). }
\label{fg:slices} 
\end{figure*}

The low-energy (1 to 6 meV) inelastic neutron scattering measurements were performed on the Multi-Axis Crystal Spectrometer (MACS)\cite{macs08} at the NIST Center for Neutron Research (NCNR).  We used a fixed final energy of 5~meV, with Be filters after the sample, and horizontal collimations of $100'$-open-S-$90'$-open, where S $=$ sample.  The middle-energy (6 to 12 meV) data were collected on the triple-axis spectrometer BT-7 at NCNR \cite{lynn12}. There we used horizontal collimations of open-$80'$-S-$50'$-$50'$  with fixed final energy of 14.7~meV and two pyrolytic graphite filters after the sample.  The high-energy (10 to 110 meV) experiments were performed on the SEQUOIA time-of-flight spectrometer at the Spallation Neutron Source (SNS), Oak Ridge National Laboratory \cite{sequoia10}.  Incident energies of 50, 100, and 180 meV were used to measure excitations from 10--34 meV, 35--70 meV, and above 70 meV, respectively.

Constant energy slices through the dynamical structure factor $S({\bf Q},\omega)$ are shown in Fig.~\ref{fg:slices}.  The  wave vectors {\bf Q} are specified in reciprocal lattice units (rlu), $(a^*, b^*, c^*) = (2\pi/a, 2\pi/b, 2\pi/c)$, where the lattice constants are $a \approx b = 3.79$~\AA, and $c\approx 13.2$~\AA.  $S({\bf Q},\omega)$ is the Fourier transform of the spin-spin correlation function.  To extract it from the measured scattering intensity, it was necessary to divide out the square of the magnetic form factor \cite{xu13}.  We used a recent determination of the Cu form factor that takes account of hybridization \cite{walt09}.  To put the scattering data in absolute units, the BT7 data were normalized to measurements of incoherent elastic scattering from the sample \cite{xu13}.  There the integration of the magnetic peaks was evaluated from scans along $(H,0.5,0)$, taking account of the calculated spectrometer resolution along $K$ of 0.087~rlu. The SEQUOIA (MACS) data were cross normalized with the BT7 data through integrated magnetic peak intensities at 32~K  and 10~meV (6~meV).  Examples of line cuts comparing the data and fits are given in \footnote{See Supplemental Material at [URL to be inserted] for line cuts of low-energy data and a plot of the effective dispersion at low energy.}.

\begin{figure}[t]
\centerline{\includegraphics[width=\columnwidth]{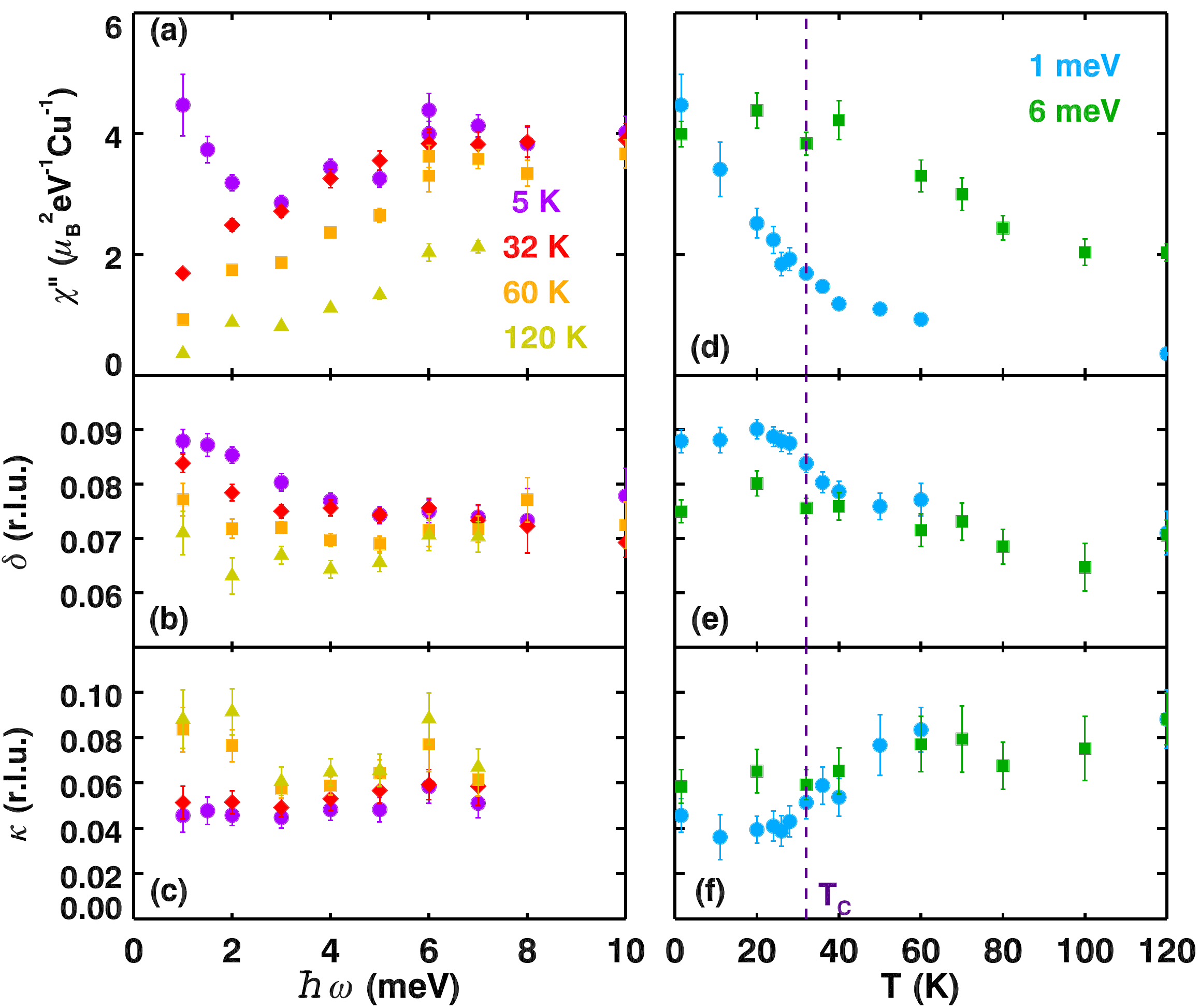}}
\caption{(color online)
Summary of fitted parameters: (a) $\chi''(\omega)$, (b) $\delta$, and (c) $\kappa$ vs.\ $\hbar\omega$, for temperatures of $\le5$~K (violet circles), 32~K (red diamonds), 60~K (orange squares), 120~K (yellow triangles);   (d) $\chi''(\omega)$, (e) $\delta$, and (f) $\kappa$ vs.\ $T$ for energies of 1~meV (blue circles), 6~meV (green squares).  Dashed line indicates $T_c=32$~K.  
}
\label{fg:sum} 
\end{figure}

The constant-energy slices in Fig.~\ref{fg:slices}(a)--(f) illustrate the dispersion of the magnetic excitations.  At low energy, we see incommensurate peaks at positions $(0.5\pm\delta,0.5)$ and $(0.5,0.5\pm\delta)$.  With increasing energy, they disperse inwards towards ${\bf Q}_{\rm AF}$ near 35~meV, and then outwards again at higher energies, following the common hour-glass dispersion \cite{fuji12a}. From slices such as those in Fig.~\ref{fg:slices}(g)--(i), one can see that the peak positions $\delta$ and widths $\kappa$ (full width at half maximum) change with temperature.  To parametrize the data, we have performed least-squares fitting with four symmetrically-positioned, normalized Gaussian peaks.  We have expressed the amplitude in terms of the imaginary part of the dynamical spin susceptibility, given by \cite{xu13}
\begin{equation}
     \chi''({\bf Q},\omega) = g^2\mu_{\rm B}^2\frac{\pi}{\hbar}\big(1-e^{-\hbar\omega/k_{\rm B}T}\big)  S({\bf Q},\omega),
     \label{eq:S}
\end{equation}
where $g\approx2$ is the gyromagnetic factor.  Since we use normalized Gaussians, the fitted amplitude parameter can be expressed in terms of the {\bf Q}-integrated local susceptibility, $\chi''(\omega)$, which, at low temperature, is essentially the magnetic spectral weight.  Examples of fits are shown in Fig.~\ref{fg:slices}(j)--(l); the results for the fitted parameters are summarized in Fig.~\ref{fg:sum}.  

Focusing on excitations below 10 meV, one can see in Fig.~\ref{fg:sum}(a) that cooling leads to an enhancement of $\chi''(\omega)$ that saturates by $T_c$---except for $\hbar\omega<3$~meV.  At $T\le5$~K, $\chi''(\omega)$ exhibits a quasi-elastic peak associated with spin-stripe correlations \cite{huck11}.  For the quasi-elastic energies, there is also a temperature-dependent shift in the incommensurability $\delta$, as shown in Fig.~\ref{fg:sum}(b).  Near $T_c$, we find $\delta\approx0.075$ rlu, but at 5~K there is a substantial upward shift towards 0.09 for $\hbar\omega < 3$~meV, effectively dispersing \cite{Note1} towards the elastic peak centered at $\delta=0.105$~rlu, which develops below $\sim T_c$ \cite{huck11}.

A complementary picture is given by the temperature dependence of $\delta$ for $\hbar\omega=1$~meV, as shown in Fig.~\ref{fg:sum}(e).  The strong shift in $\delta$ begins slightly above $T_c$, at $\sim40$~K, which corresponds with the onset of strong superconducting correlations \cite{steg13}.  
The growth of $\chi''(1{\rm\ meV})$ also takes off below 40~K, and the line width $\kappa$ decreases.

The magnetic incommensurability is a consequence of the charge carriers forming intertwined stripes \cite{frad14}.  Changes in the stripe spacing with temperature are reflected in $\delta$ and are tied to the behavior of the charge carriers; for example, in La$_{1.875}$Ba$_{0.125}$CuO$_4$ there is an abrupt jump in $\delta$ at a structural transition associated with pinning of the charge stripes \cite{fuji04}. In the present case, where there is also a structural transition \cite{wen12a}, the quasi-static stripe correlations appear to develop in a cooperative fashion with the superconductivity.

\begin{figure}[t]
\centerline{\includegraphics[width=0.85\columnwidth]{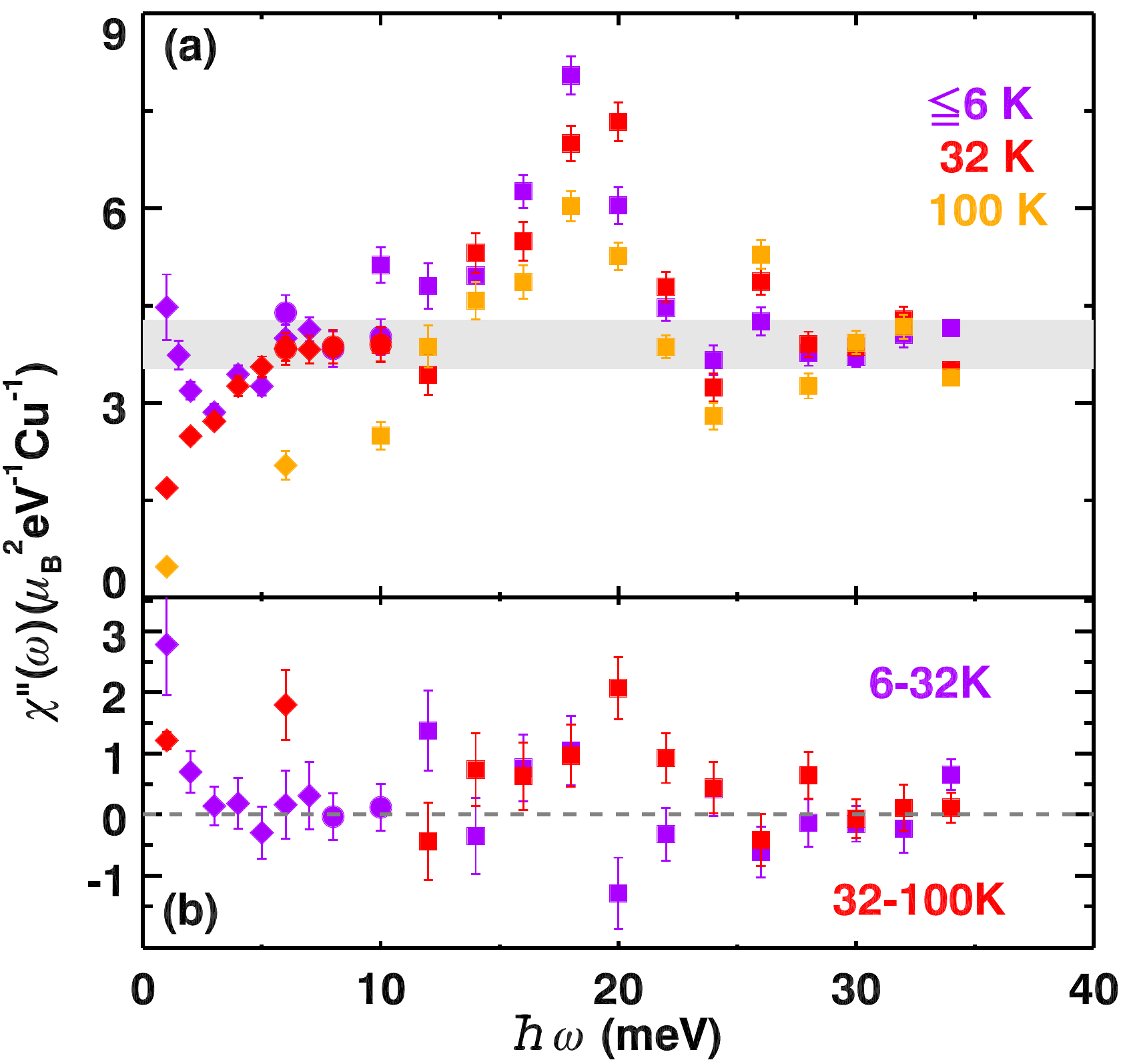}}
\caption{(color online)
(a) Wave-vector-integrated local susceptibility $\chi''(\omega)$ measured at temperatures of $\le 6$~K (violet), 32~K (red), and 100~K (orange).  (b) Difference in $\chi''(\omega)$ between $\sim6$ and 32~K (violet), and between 32 and 100~K (red).  In (a) and (b), diamonds were measured on MACS (base $T=1.5$~K); circles on BT7 (base $T=5$~K); squares on SEQUOIA (base $T=6$~K). }
\label{fg:chi} 
\end{figure}

Figure~\ref{fg:chi} presents the results for $\chi''(\omega)$ over a broader energy range at select temperatures above, below, and at $T_c$.   In optimally- and over-doped cuprates, the spin gap for $T<T_c$ is generally observed to be comparable to the superconducting gap $\Delta$, with a resonance peak appearing at $E_r\approx 1.3\Delta$ \cite{sidi04,yu09}.  Measurements such as Andreev reflection indicate that $\Delta\approx 2.5k_{\rm B}T_c$ \cite{deut99,gonn01}, yielding a prediction of $\Delta\approx 7$~meV for our sample, consistent with the gap measured on the Fermi arc by angle-resolved photoemission \cite{he09}.  Correspondingly, one predicts $E_r\sim 9$~meV.

It is quite clear from Fig.~\ref{fg:chi} that there is no meaningful spin gap for $T<T_c$ on the predicted scale of 7 meV.  While there is a depression of $\chi''(\omega)$ below 6 meV at $T_c$, it should be noted that $\chi''$ must decrease to zero at $\hbar\omega=0$.  There is a definite enhancement of the signal below 3~meV at low temperature.  Similarly, there is no obvious resonance feature.  While $\chi''(\omega)$ does exhibit a peak near 18 meV, that peak is already present at 100~K, and there is no significant correlation between the peak intensity and the development of superconductivity.  The conclusion of an independent neutron scattering study is that the peak is a consequence of spin-phonon hybridization \cite{wagm14}.

To put the absolute magnitude of $\chi''(\omega)$ in context, we can compare with the signal from spin waves in La$_2$CuO$_4$ \cite{cold01,head10}.  Using the results of spin-wave theory \cite{mano91}, we find that $\chi''(\omega) = (Z_\chi/J_{\rm eff})\,\mu_{\rm B}^2/{\rm Cu}$, where $Z_\chi\approx0.5$.  From experiment, the value of the effective superexchange energy, $J_{\rm eff}$, describing the low-energy dispersion is 128 meV \cite{cold01,head10}.  From this we obtain $\chi''(\omega)=3.9\,\mu^2_{\rm B}\,{\rm eV}^{-1}{\rm Cu}^{-1}$, which is indicated by the gray bar in Fig.~\ref{fg:chi}.   It is strikingly similar to the magnetic spectral weight found for our sample of \lbco\ with $x=0.095$.   The degree of similarity may be coincidental, as parameter and data normalization uncertainties are on the order of 20\%.  The point is that the strong magnetic response cannot come from only a small fraction of the sample.  In combination with the evidence for bulk superconductivity \cite{home12}, it appears inescapable that superconductivity and antiferromagnetic spin correlations must coexist locally. 

From the perspective of competing orders \cite{deml01}, coexistence of superconductivity and spin-density-wave order (SDW) requires one or both these orders to be weak.  While the true SDW order is weak in our sample, the presence of the strong magnetic spectral weight at energies far below the superconducting gap, together with the optical evidence for a substantial superfluid density, is problematic.  Similarly, it would be difficult to rationalize the experimental observations in terms of disorder effects alone \cite{ande10}.  A different approach is necessary.

A way to reconcile the coexisting spin fluctuations and superconductivity is to relax the expectation of spatial uniformity.  We know from their incommensurability that the locally-antiferromagnetic spin correlations are spatially modulated with a period of roughly 9 lattice spacings.  It is also possible for the pair-wave function to be spatially modulated and phase shifted so as to minimize overlap with the low-energy spin correlations.  If the pair wave function is sinusoidally modulated with the same period as the spin correlations, so that its amplitude varies from positive to negative, then it represents a pair density wave (PDW) \cite{berg09b}; such a state has been proposed to explain the decoupling of superconducting layers in \lbco\ with $x=\frac18$ and related systems \cite{berg07,hime02}.  It is also possible to have the amplitude modulated but without a sign change, in which case it should have half the period of the spin correlations.  Recent variational calculations applied to the $t$-$J$ model have found that the energies of the PDW and the in-phase striped superconductor are very close, and both are competitive with the uniform $d$-wave state \cite{corb14}.   

Previous studies \cite{wen12,steg13} have shown that application of a strong $c$-axis magnetic field to \lbco\ with $x=0.095$ causes a decoupling of the superconducting layers in a manner consistent with the PDW scenario for the $x=\frac18$ composition in zero field.  Given that the PDW state is quite sensitive to disorder \cite{berg09b}, the robust superconductivity found for $x=0.095$ in zero field may favor an in-phase striped superconductor.  

As already noted, gapless spin fluctuations have also been detected by neutron scattering in underdoped \lsco\ \cite{lee00,chan07,lips09,kofu09}.   In the case of La$_{1.875}$Sr$_{0.125}$CuO$_4$, where charge stripe order has recently been reported \cite{chri14,tham14,crof14},  Kofu {\it et al.} \cite{kofu09} proposed that the spin excitations below 4 meV come from different spatial regions than the excitations above 4 meV, thus invoking large scale phase separation to maintain consistency with the spin-gap paradigm.  We actually share the concept of phase separation, but on a much shorter length scale.  We argue that the low-energy spin excitations come from the spin stripes coexisting with the superconductivity.  Regarding the possibility of large-scale phase separation in \lsco, we note that, 
for superconducting samples with $0.06<x<0.10$, it has been concluded from muon spin rotation ($\mu$SR) studies that there is static, inhomogeneous magnetic order throughout the volume at $T<1\ {\rm K}<<T_c$, with any non-magnetic regions being smaller in size than 20~\AA\  \cite{nied98}.  These materials are also believed to be bulk superconductors, which again is consistent  with intertwined coexistence.

To rationalize the differences between superconducting samples with and without spin gaps, we suggest the following scenario.  At optimal doping and above, where the pair wave function is spatially uniform, it is favorable to gap out any residual spin fluctuations at $\hbar\omega<\Delta$. At lower doping,  when strong low-energy spin-stripe correlations are present in the normal state, it may be too energetically costly to gap the spin excitations.  Instead, it may be favorable to modulate the pair wave function to avoid the antiferromagnetic spin correlations by intertwining with them \cite{hime02,berg09b}.  This scenario is consistent with the idea that antiferromagnetism and superconductivity are closely associated \cite{scal12a,norm11}, but it suggests the need for a pairing mechanism \cite{emer97} that goes beyond  the conventional concept of ``pairing glue'' \cite{scal12a}.

Finally, we note that spin-gapless superconductivity is not limited to ``214'' cuprates.   $\mu$SR measurements also indicate static magnetic fields in superconducting Y$_{1-x}$Ca$_x$Ba$_2$Cu$_3$O$_{6.02}$ for hole concentrations similar to those in \lsco\ \cite{nied98}.  Furthermore, neutron and $\mu$SR results for superconducting \ybco\ with hole concentrations $p\lesssim0.08$ indicate coexisting gapless spin correlations \cite{hink08,stoc08,haug10}.  More generally there have been theoretical and experimental papers proposing the relevance of a PDW state to understanding the pseudogap in cuprates such as \ybco, especially at high magnetic field and low temperature \cite{lee14,yu14}.

We are grateful for helpful comments from E. Fradkin, S. A. Kivelson, and T. M. Rice.  Work at Brookhaven was supported by the Office of Basic Energy Sciences (BES), Division of Materials Sciences and Engineering, U.S. Department of Energy (DOE), through Contract No.\ DE-AC02-98CH10886.   This work utilized facilities at the NCNR supported in part by the National Science Foundation under Agreement No. DMR-0944772.  The experiments at ORNL's SNS were sponsored by the Scientific User Facilities Division, BES, U.S. DOE.


%

\renewcommand{\thefigure}{S\arabic{figure}}
\setcounter{figure}{0}
%

%
%
\vspace{20pt}

The constant-energy slices of the inelastic neutron scattering data for La$_{1.905}$Ba$_{0.095}$CuO$_4$, examples of which are shown in Fig.~1 of the main paper, were fit as a function of wave vector $(H,K)$.  This procedure makes good use of the available data; however, two-dimensional representations of the data and fits can be difficult to compare at a detailed level.  Here we present some representative line cuts that allow a direct comparison of the data and the fits.  Figure S1 shows cuts along $(H,0.5)$ and $(0.5,K)$, corresponding to sample temperatures of 1.5~K and 32~K.  Each slice has a with of 0.02 rlu in the transverse direction.  Note that the range over which the actual fitting was performed was within 0.2 rlu of ${\bf Q} = (0.5,0.5)$.

\begin{figure*}[t]
\centerline{\includegraphics[width=\columnwidth]{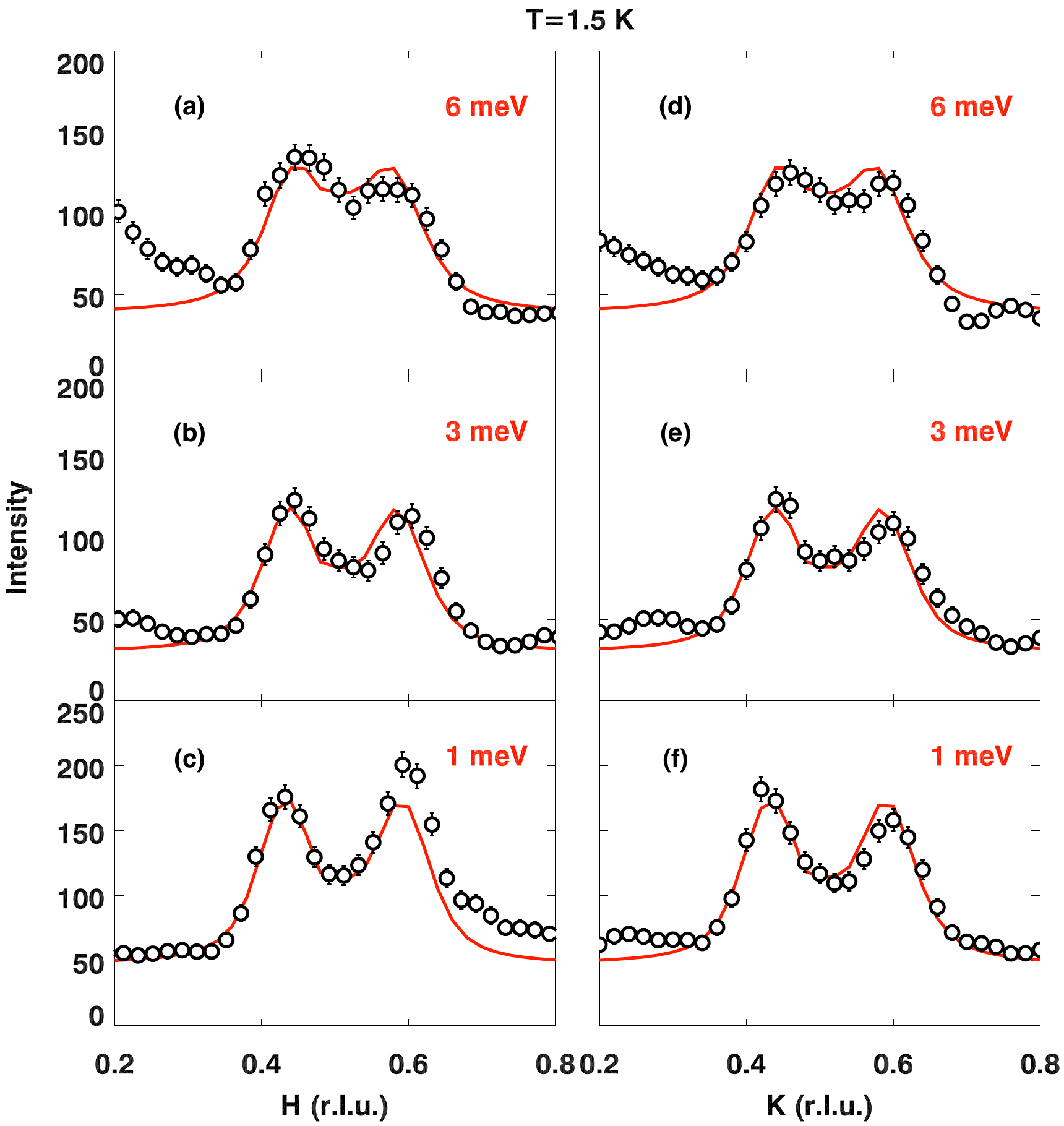}
\hfil \includegraphics[width=\columnwidth]{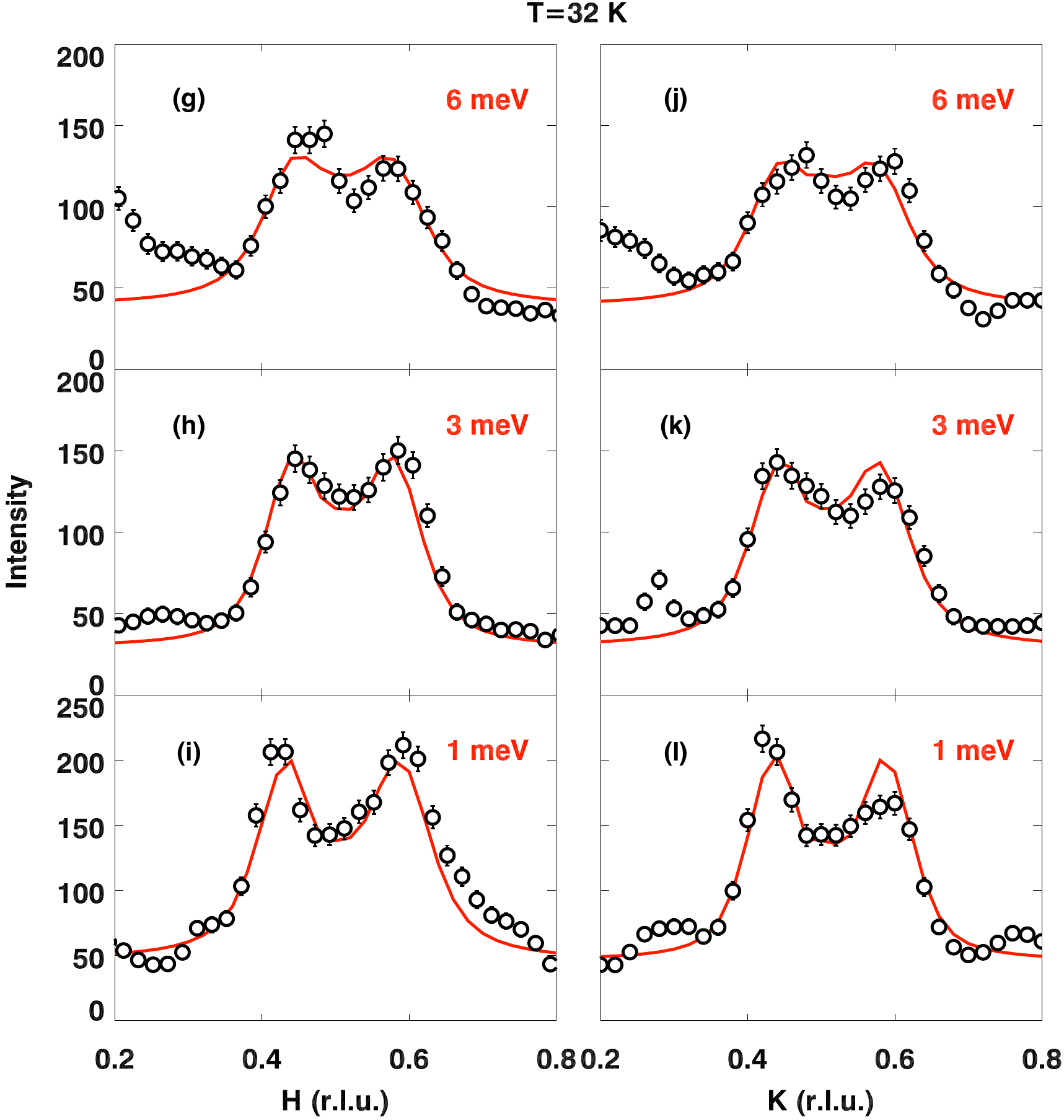}}
\caption{(color online)
Plots of line cuts measured at (a)--(f) $T=1.5$~K, (g)--(l) 32~K.  Panels (a)--(c) and (g)--(i) correspond to cuts along ${\bf Q} = (H,0.5)$; (d)--(f) and (j)--(l) are along $(0.5,K)$.  The excitation energies are 1 meV for (c), (f), (i), (l); 3 meV for (b), (e), (h), (k); 6 meV for (a), (d), (g), (j).  Symbols represent data, lines indicate the fitted function. }
\label{fg:cut1} 
\end{figure*}

In Fig.~S2, we present data from Fig.~2(b) in the form of an effective dispersion of the magnetic excitations.  The rapid inward dispersion of the excitations at low energy, followed by a vertical rise is quite unusual.  We do not know of any models that directly describe such behavior.  Note that the unusual dispersion occurs below 4 meV, the same energy range in which Kofu {\it et al.} [10] observed distinct behavior in the superconducting phase of La$_{1.875}$Sr$_{0.125}$CuO$_4$.

\begin{figure}[t]
\centerline{\includegraphics[width=\columnwidth]{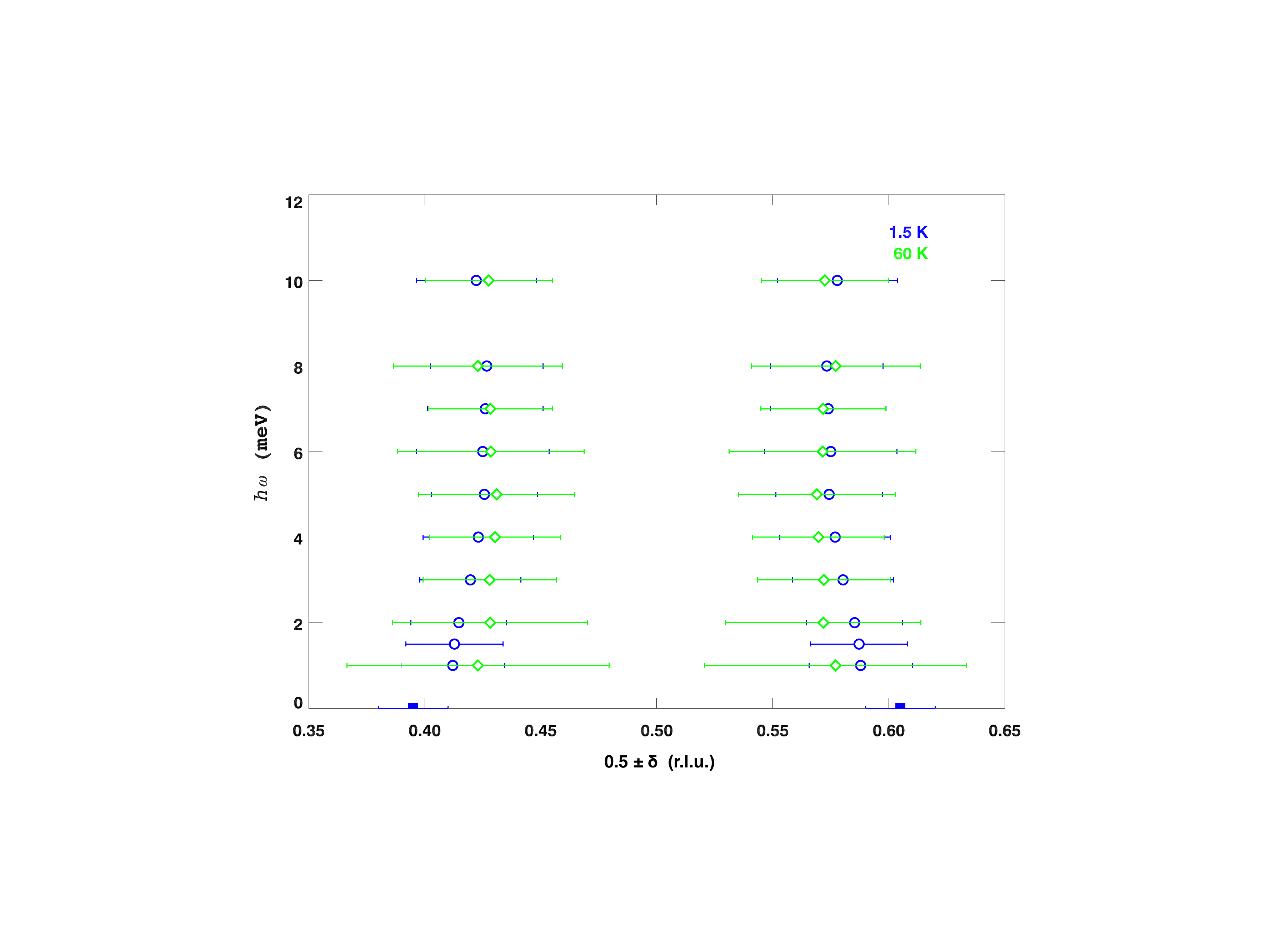}}
\caption{(color online)
Plot of the effective dispersion of magnetic excitations, $E$ vs.\ $0.5\pm\delta$.  Blue circles (green diamonds) correspond to $T=1.5$K (60 K).  Blue squares represent the elastic magnetic peak positions at low temperature from a previous study [21].  Horizontal bars indicate peak widths. }
\label{fg:disp} 
\end{figure}

\end{document}